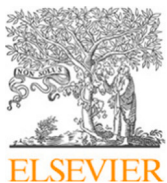
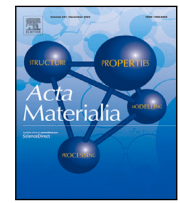

Full length article

# Effect of surface orientation on blistering of copper under high fluence keV hydrogen ion irradiation

A. Lopez-Cazalilla [a,*], C. Serafim [a,b], J. Kimari [a,c], M. Ghaemi [a], A.T. Perez-Fontenla [b], S. Calatroni [b], A. Grudiev [b], W. Wuensch [b], F. Djurabekova [a]

[a] *Department of Physics, University of Helsinki, P.O. Box 43, FI-00014, Finland*
[b] *CERN, European Organization for Nuclear Research, 1211 Geneva 23, Switzerland*
[c] *KTH Royal Institute of Technology, Nuclear Engineering, Stockholm SE-106 91, Sweden*

A B S T R A C T

Copper and hydrogen are among the most common elements that are widely used in industrial and fundamental research applications. Copper surfaces are often exposed to hydrogen in the form of charged ions. The hydrogen ions can accelerate towards the surface, resulting in an accumulation of hydrogen below the surface. Harmless in low concentrations, prolonged hydrogen exposure can lead to dramatic changes on copper surfaces. This effect is visible to the naked eye in the form of blisters densely covering the exposed surface. Blisters are structural modifications that can affect the physical properties of the surface including, for example, vacuum dielectric strength.

Using scanning electron microscopy we found that the blistering of the irradiated polycrystalline copper surface does not grow uniformly with ion fluence. Initially, only some grains exhibit blisters, while others remain intact. Our experiments indicate that grains with the {1 0 0} orientation are the most prone to blistering, while the grains oriented in the {1 1 0} are the most resistant to it. Moreover, we noticed that blisters assume different shapes correlating with specific grain orientation.

Good agreement of experiments with the atomistic simulations explains the difference in the shapes of the blisters by specific behavior of dislocations within the FCC crystal structure. Moreover, our simulations reveal the correlation of the delay in blister formation on surfaces with certain orientations compared to the others with the dependence of the hydrogen penetration depth and the depth and amount of vacancies in copper on the orientation of the irradiated surface.

## 1. Introduction

The interaction of hydrogen with metals has attracted much attention recently because of various energy technology solutions, such as energy storage, future nuclear fusion power plants, hydrogen fuel cells and oil refining [1–4]. One of the major challenges in the use of hydrogen is hydrogen-induced degradation, corrosion, cracking and blistering [5,6]. Similarly, in technological applications related to particle accelerators or fusion power plants, metal surfaces can also be extensively exposed to hydrogen in the form of protons or negatively ionized atoms. One of the most detrimental effects that metal surfaces can experience under these conditions is blistering. The mechanisms behind blistering have been extensively studied in many materials to prevent malfunction or failure of devices exposed to hydrogen or noble gas ions [7]. These light species create defects in the lattice that can accumulate at grain boundaries or other lattice imperfections [8–10].

Under hydrogen ion irradiation, vacancies formed in atomic collisions with the bulk lattice atoms are capable of accommodating more than a single hydrogen atom [11–13]. Accumulation of multiple hydrogen atoms within a vacancy leads to expansion of the vacancy into a larger vacancy cluster that can accommodate even more hydrogen atoms [14], eventually resulting in stable bubbles at least 1.5 nm in size [15] and larger. On the surface, these bubbles are observed in the form of blisters [16–20].

Blistering is a phenomenon that has been observed in many metals regardless of their crystalline structure, body-centered cubic (BCC) or face-centered cubic (FCC). For example, tungsten, which is one of the most studied BCC metals because of its application for plasma-facing components in future nuclear fusion power plants, remains susceptible to the adverse effect of blistering [21–24] despite its otherwise superior properties, such as high melting point, good thermal conductivity, low thermal expansion, and high strength at high temperatures. Moreover, this effect has been reported in many other BCC metals such as V [25], Nb [26], carbon steel [27–29] and Fe [30–32]. FCC metals, such as Al [6,33–38] and Cu [18–20,39–42], experience blistering under different conditions of temperature, ion energy and fluence. For increasing the durability of these materials, hence improving their performance, it is important to understand the mechanisms of blistering in metals in order to mitigate its effects, since the properties of these materials change when exposed to light gas species [43,44]. This effect is particularly interesting in Cu, a metal that has a wide range of applications, including natural gas pipelines, the electronics industry, architecture and the storage of nuclear fuel waste [45–47]. It is also one of the most widely used metals in accelerator technologies because of its high electrical and thermal conductivity, and mechanical properties [48]. More precisely, Cu is used in radio frequency quadrupoles (RFQ) which act as injectors of protons from the source in the Large Hadron Collider (LHC) [49] and in neutral beam injectors in the International Thermonuclear Experimental Reactor (ITER) [50]. A proper understanding of blistering is crucial to avoid the frequent replacement of expensive components and disruption of regular operation of these devices.

Surface modifications in the form of protrusions can be seen in a microscope on metal surfaces. In $H_2$ atmosphere the protrusions were observed on both Al [38] and Ni [51] surfaces. Moreover, the shapes of these protrusions on Al surface were found to be different on the grains with different orientations, when the surface was exposed to low energy and high fluences molecular $H_2$ plasma [52]. These protrusions were shown to have square, circular and triangular shapes on the {1 0 0}, {1 1 0} and {1 1 1} surfaces, respectively. The conditions of exposure to hydrogen beams, i.e. energy and fluence, may play a role in the final surface modification. Some discussion related to plastic deformations in FCC lattices due to hydrogen accumulation under the surface is offered to explain the shapes of the protrusions. However, the proposed model does not provide sufficient atomic level insight into the effect of crystallographic orientation at any stage of protrusion development in these materials.

Computational methods, such as molecular dynamics (MD) [53], which simulate explicitly the dynamics of atomic motion under different external conditions, can provide crucial insights on mechanisms of plastic deformations as well as explain the dependence of the studied processes on crystallographic orientations of the material structure. For instance, valuable insights on the mechanisms of small gas bubbles growth in abundance of He and H atoms were obtained in Ni [54] and Cu [42,55]. However, the bubble growth near the surface which can result in surface blisters in FCC metals has not been yet addressed with an atomistic level detail.

Simulating the entire process of blister formation at the surface with MD methods is infeasible due to the size and time scale limitations. Generally, the binary collision approximation (BCA) methods are considered to fit better for the studies of implantation of light ions into a target [56–58]. The information about the ion depth profile and vacancy concentration, which is easily available from these calculations, can be used to model the size of forming blisters. Nevertheless, these models do not consider the crystal orientation as a parameter important for the blistering process, although it was shown to be significant for Ni [51] and Al [38,52]. For Ni, a lower density of blisters was observed for (1 0 0) grains compared to (1 1 1) ones. Similarly, for Al pillars [38], larger density of blisters were observed for (1 1 1) planes than in (1 0 0), and no blister was found in (1 1 0) planes of the pillars. Some authors [59,60] studied the effect of the crystal orientation on the penetration of He in Cu for several energies in the range of tens of keV. Moreno and Eliezer [59] have reported the observed dependence of the size of blisters on surface orientation, attributing this dependence to the different He depth profiles in different directions in Cu and CuBe. Unfortunately, the reported difference in He depth profiles is insufficient to explain the experimental results in the present work.

To explain blister formation in W [61–63], Au [64] and Ni [65], the defect formation (in terms of displacements per atom) was considered in combination with the depth profile of gaseous ions of different incident energies. Some details on bubble growth mechanisms were obtained in Fe [66] and Al [67], highlighting the importance of the interaction of H with the created vacancies. However, in these studies, the crystal orientation of the matrix as well as the effect of the surface were not included.

In the present study, the blisters were observed in radio-frequency quadrupole (RFQ) structures, which are the first accelerator element downstream the $H^-$ ion source in the LHC injector chain at CERN. Here the stray beam of the $H^-$ ions impacts the copper surfaces at the RFQ entrance with energies between 45 keV and 3 MeV. During the LHC runs, the surface is exposed to the beam for very long time, which becomes sufficient to cause severe deformation of the surface. To study this process, we used a combined experimental and theoretical approach to explain the non-uniformity of shapes and sizes of blisters covering the polycrystalline copper surface under prolonged exposure to $H^-$ ion beam of 45 keV energy.

## 2. Methods

### 2.1. Experimental methods

In our experiments, we used Cu-OFE (oxygen-free electronic-grade copper, UNS C10100) electrodes that were prepared by diamond machining from a multi-directional forged bar [68]. The shape of the electrodes used in this work is compatible with the large electrode system described in detail in Ref. [42,69]. After machining, the degreased electrodes underwent the thermal treatment (835 °C, with the annealing ramp of 100 °C/h) applied for 20 min. This imitates the vacuum brazing cycle employed in the fabrication of the actual RFQ.

The irradiation of the sample was performed at CERN using an identical twin $H^-$ source that injects into the operational RFQ. The sample was installed in a vacuum downstream of the source with a fixture, permitting beam incidence perpendicular to the sample surface. Beam energy was 45 keV at a fluence of $\sim 1.3 \times 10^{19}\,H^-/cm^2$ with an exposure time of 40 h, and a duty cycle of about 0.1% at a repetition rate of 1 Hz.

According to Ref. [70], $H^-$ is fully ionized into $H^+$ within less than 1 atomic diameter after entering the copper surface, and calculations show that $H^+$ at 45 keV has a Bragg peak at a depth of about 300 nm.

Prior to irradiation, the sample was analyzed using electron backscatter diffraction (EBSD), where the orientation of the different grains can be detected with a certain range of accuracy. This allows us to follow the effect of the incoming irradiation in the differently oriented grain, and obtain information from it. For more information see Figure S1 in Supplementary material Section A.1.

Blisters on the sample were imaged by using a Scanning Electron Microscope (SEM) model Sigma from Zeiss. Additionally, Focused Ion Beam (FIB) milling was performed to obtain the cross-sectional view of the grains using a Pt protective layer.

### 2.2. Computational methods

In our computational studies, we performed MD simulations using the PARCAS code [53]. We used the embedded atom model (EAM) interatomic potential, developed by Mishin et al. [71] (*Mishin* potential) to describe the Cu–Cu interactions. We also used an EAM potential for the H–H interactions, following the works in Refs. [72–74], while the purely repulsive ZBL potential [75] was used for the Cu–H interactions. The use of a repulsive potential is motivated by the low solubility limit of H in Cu [76]. Moreover, in this approximation, we were able to generate sufficient pressure that is exerted at the walls of the void within an MD cell [42], while missing the occasional appearance of H atoms in interstitial positions within the Cu lattice.

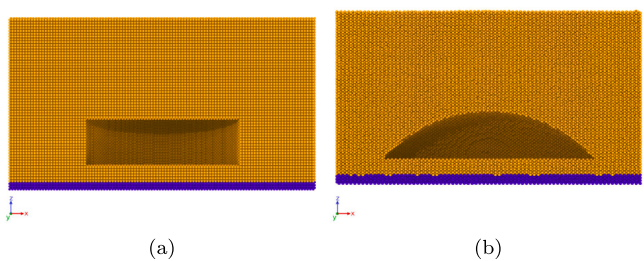

**Fig. 1.** Slice of the different simulation cells. In blue the fixed layer at the bottom. (a) Disk-shaped void and (b) hemisphere-shaped void. (For interpretation of the references to color in this figure legend, the reader is referred to the web version of this article.)

All cells in our simulations were of the same size, which was ∼30 nm in the $x$–$y$ directions and 15 nm in the $z$-direction, containing ∼1 040 000 atoms. The generated cell was relaxed at 600 K and 0 kbar (NPT) [77] with periodic boundary conditions in all directions for 50 ps. The main simulations of this study were performed using periodic boundary conditions only in the $x$-$y$ direction, while the cell in the $z$-direction had an open surface at its top. The cell was rotated to align the $z$-axis with one of the low-index crystallographic directions {1 0 0}, {1 1 0} and {1 1 1} to imitate differently oriented grains. At the bottom of the cell, a fixed atomic layer prevented the cell from moving during the simulations (see Fig. 1, blue slabs). After the surface was opened, the cells in all three orientations were additionally relaxed for 100 ps at 600 K in NVT using the Berendsen thermostat [77]. The same temperature using the Berendsen thermostat was applied in all the remaining simulations, which were performed for 100 ps of the simulated time.

In the prepared cells of three orientations, we inserted two types of voids in the shape of a disk and a hemisphere (see Fig. 1). These shapes were selected to imitate the shapes found in the SEM images of the bubbles grown in Cu during the extensive H$^-$ irradiation (see Fig. 3). Moreover, the chosen geometries can enhance the interaction of the void with the surface to compensate for the limited timescale of MD simulations. For consistency, we kept the volume of all the created voids the same (∼7 × 10$^3$ nm$^3$) regardless of their shapes. All bubbles were positioned 9 nm below the surface measured from the top of the bubble.

The pressure within the bubbles was controlled by the concentration of H atoms in terms of $n_{H/Vac}$ (the number of H atoms per missing Cu atom). The H atoms are randomly distributed within the void at a distance not shorter than 0.09 nm, which is larger than the bond length predicted by the H–H potential [74]. Although we simulated the bubbles with the different concentrations of the H atoms, we were able to observe structural modifications only under the pressure of 20 kbar ($n_{H/Vac} \approx 1.2$). To ensure the sensible result within 100 ps, we increased the H concentration to $n_{H/Vac} = 2$ (30 kbar) as in [42], while performing only one test simulation with $n_{H/Vac} = 1.2$ for the disk-shaped bubble, which was located closer to the surface than other bubbles (see Supplementary material Section A.2). We also show in the Supplementary material Section A.3 that under the pressures which we use in the current simulations, the likelihood of formation of H$_2$ molecules is low, while in experiments, under the lower pressures and longer time scales the gas may consist of molecules. We observe how the hydrogen contained in the bubble is in atomic state, being this pressure enough to not allow the formation of H–H bonds.

The solubility of H in Cu, at similar conditions to these reproduced in our computational study (600 K and 20 kbar) is very low [76]. Hence we can assume that the H atoms present in the Cu lattice accumulate in a void, which allows us to focus on the response of the material to the hydrostatic pressure of the hydrogen gas within the bubble.

For visualization of the results, we used the Open Visualization Tool (OVITO) [78].

To estimate the hydrogen depth profile for different crystallographic orientations, we additionally used the MDRANGE code [79], which is an MD code efficiently calculating ion range profiles and their lateral stragglings within the recoil interaction approximation (RIA). For instance, the code was successfully applied to calculate the channeling maps in materials with different crystal structures [80]. In this approximation, solely the forces acting between the ion and the target atoms are considered, disregarding the further evolution of the cascades that involve interactions amongst the target atoms. Statistically significant results were obtained by running 10 000 simulations of 45 keV H impacts for each low-index crystallographic orientation of the copper surface. Since these simulations are limited to the interactions of the impacting ions only, it is not possible to obtain directly the number of produced vacancies to be compared between the different crystallographic directions. Both distributions of the produced defects and the ion range were obtained with the help of a more conventional binary collision approximation approach, using the CASWIN code [81] simulating 50 000 impacts of 45 keV. Although the crystal structure is not taken into account in these simulations, we use them to correlate the vacancy and the primary recoils depth distributions generated by the passing hydrogen ions in Cu. More detail on the CASWIN simulations can be found in Supplementary material A.4.

## 3. Results and discussion

In this section when discussing the experimental results, we will refer as blister to those bubbles of large size (larger than hundreds of nm) and not entirely rounded, that are formed close to the surface and leave track at the surface. If the bubbles are rounded and the size smaller than the called blisters, we will simply use bubble. However, in the computational results, we will use the term protrusion to name this effect at the surface, since the scale of the computational results differ from the experimental ones.

### 3.1. Blister shapes and sizes on polycrystalline cu surface under different h ion fluence

The orientation of the grains on the Cu surface was determined before irradiation using the EBSD technique. The full map of the analyzed region can be seen in Supplementary Figure S1. We see the overall prevalence of the blue, green and red colors on the map, which indicate the preferential orientation of the grains is in one of the low-index directions, such as [1 1 1] (blue), [1 1 0] (green) and [1 0 0] (red). To verify the effect of the H$^-$ exposure on the bubble growth in the sample, we selected two different areas with varying exposures to the ion beam, i.e. at different distances from the center of the beam. These areas are shown in Figure S1: in blue a transition area between the core of the beam and the halo, and in magenta a high fluence irradiation area corresponding to the beam core.

Fig. 2 shows a section of the transition area of the Cu surface exposed to the H$^-$ ion irradiation for 40 h reaching the fluence of ∼ 1.3 × 10$^{19}$ cm$^{-2}$. Here we selected the regions within the transition area (i.e. the low fluence) across two grains with notably different orientations ((1 1 1)/(1 0 0) in Fig. 2a and (1 0 0)/(1 1 0) in Fig. 2c according to the EBSD map). In Fig. 2a we observe the formation of dimples in the center of several blisters. The appearance of dimples in these blisters can be due to a larger hydrogen pressure inside them, inducing a stronger plastic deformation at the center over the surface, where the stress is higher. This causes a more deformed region of the blister at the top. A FIB-SEM from Zeiss was used to mill the sample to obtain the cross-sectional views of the two grains in question. Visualizing the sample, we systematically observe a thin black layer which corresponds to carbon, and it is observed in all the grains considered (see for instance Fig. 2d). The carbon layer results from the surface cracking of hydrocarbons in the residual gas of the vacuum of our irradiation facility. This phenomenon is known to occur upon electron or ion surface irradiation also in high vacuum environments, such as scanning electron microscopes or plasma fusion experiments. In Fig. 2b, the blue and red squares outline the regions that are zoomed in Figs. 2d and 2e, respectively, while the green line in Fig. 2c indicates the area where the cross-sectional view was obtained for the (1 1 0) grain found on the same surface as the grains shown in Fig. 2a (for a detailed location of the grains see EBSD map in Figure S1a).

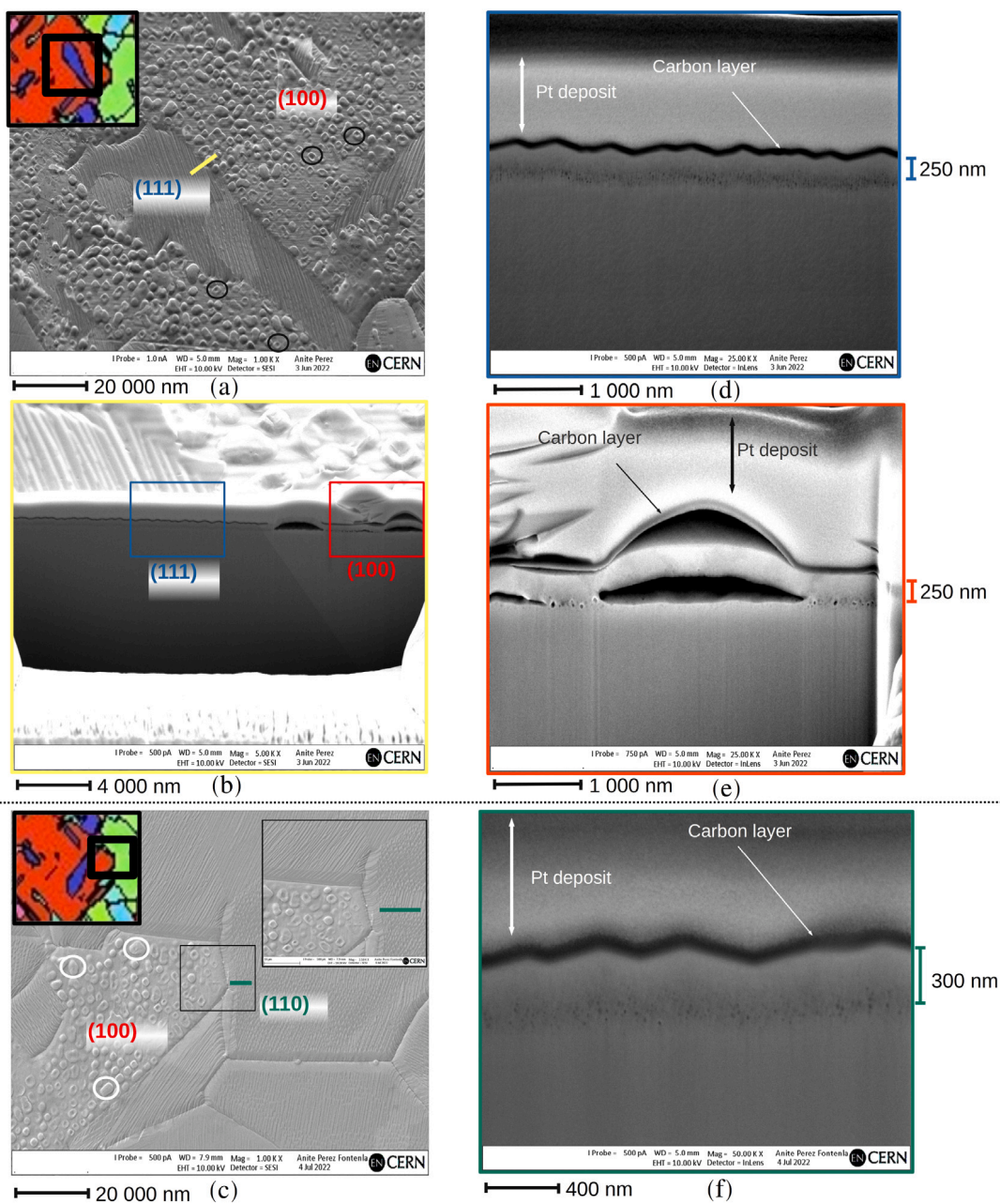

**Fig. 2.** SEM images of the Cu surface irradiated by H$^-$ ions up to a dose of $\sim 1.3 \times 10^{19}$ H$^-$/cm$^2$ at the transition area. (a) The surface of two grains is marked as (1 1 1) and (1 0 0). A yellow line marks where the cross-section was carried out. Black circles mark some of the dimples. (b) Cross-sections of the two grains (1 1 1) and (1 0 0) (corresponding to the yellow line in (a)), shown in Figures (d–e) respectively. (c) Surface of the two grains marked as (1 0 0) and (1 1 0). Inset: Zoom of the boundary region between the two grains. (f) Cross-section of a (1 1 0) grain in the transition area shown in (c) by a green line. A Zeiss Crossbeam 540 Focused Ion Beam (FIB)/Scanning Electron Microscope (SEM) was used to perform the cross-section on the sample. Imaging was performed using InLens Secondary Electron Detector at 10 keV and various magnifications. The areas shown in the cross sections are marked in red, blue and green for (1 0 0), (1 1 1) and (1 1 0), respectively. The horizontal dashed line separates the two different regions explored at the transition area. EBSD maps of the regions explored can be found in insets of (a) and (c) respectively. For a view of the complete EBSD map, see Figure S1b. (For interpretation of the references to color in this figure legend, the reader is referred to the web version of this article.)

In Fig. 2a and 2c, we clearly see that blisters of approximately square shapes are fully developed on the (1 0 0) grain, while no trace of blister formation is seen in the same images on the (1 1 1) and (1 1 0) grains. Nevertheless, in Fig. 2c we observe that close to the boundary between (1 1 1) and (1 1 0) grains, some blisters exhibit almost rhomboidal shape and the height of them is not as prominent as in other parts of the grain. Besides, we observe different shapes due to the coalescence of blisters (marked in white in Fig. 2c). In the cross-sectional image of the grain (see Fig. 2b), we see a layer of small bubbles grown at the same depth (about 250–295 nm) under the surface of both grains. However, in the (1 0 0) grain, the bubbles are more developed with some of them growing much larger than the others (see a zoom-in image in Fig. 2e), which result in big blisters on the surface. We note that the shape of the blisters is close to hemispherical, with a fairly flat bottom part. Moreover, in Fig. 2e), we observe dual blistering, which corresponds to a secondary blister formation caused by the accumulation of hydrogen between the substrate and the C layer deposited during irradiation. Surprisingly, the bubbles grown in the (1 1 1) grain are only in an initial state of nanometric size (see Fig. 2d). The bubbles under the surface in the (1 1 0)-oriented grain are grown deeper (305 and 335 nm under the surface) and even smaller (see Fig. 2e and note the different scale bar). Additionally, we observe that at the non-yielded grains a roughness pattern (i.e. faceting) is present, especially noticeable from the undulations in the carbon layer (see Fig. 2d and 2f). Faceting is a consequence of the thermal treatment on the sample, and it has been previously observed for copper [82,83]. The role of the surface patterning on the plastic deformation of the top Cu layer can influence the process of blister formation. Nevertheless, a detailed investigation into this specific effect lies beyond the scope of this work, and we will investigate this effect in the near future.

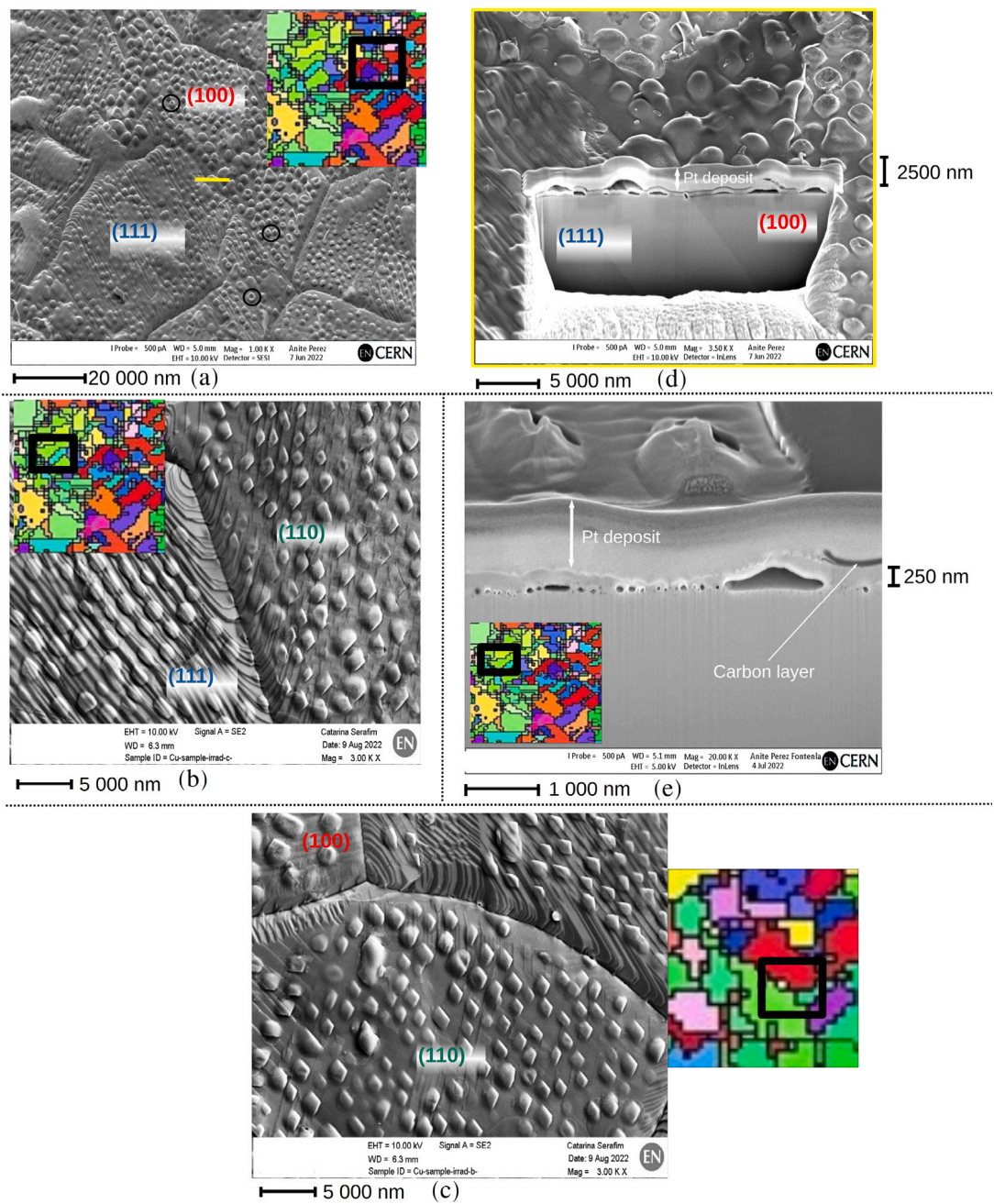

**Fig. 3.** SEM images of the Cu surface irradiated by H⁻ ions up to a dose of ∼ 1.3 × 10¹⁹ H⁻/cm² at the high fluence area. Surface of two grains explored: (a) (1 1 1) and (1 0 0), (b) (1 1 1) and (1 1 0) and (c) (1 0 0) and (1 1 0). Cross-sections of the two grains (d) (1 1 1) and (1 0 0), and (e) (1 1 0). A Zeiss Crossbeam 540 Focused Ion Beam (FIB)/Scanning Electron Microscope (SEM) was used to perform the cross section on the sample. Imaging was performed using InLens Secondary Electron Detector at 10 keV and various magnifications. Vertical and horizontal dashed lines separate the three different regions explored at the high fluence area. Black circles mark some of the dimples. EBSD maps of the regions explored can be found in insets of (a), (b) and (c) respectively. For a view of the complete EBSD map, see Figure S1c.

Fig. 3 shows the surface with different grain orientations in the region of the high irradiation fluence. The zoom-out view of this region in 3a shows that the surface, including the grains with [1 1 1] orientation, is covered by blisters quite uniformly. Again, here we observe dimples in the (1 0 0) grain, some of them resemble small holes. In the latter, the hydrogen pressure could have caused the yield of the blister at the center. Yet again, we see well-developed blisters on the (1 0 0) grain, while on the (1 1 1) grain, the blisters are weaker and developed to a lesser extent than on the (1 0 0) grain. However, a qualitative analysis of these grains shows a difference in the shapes of the blisters grown on the two explored grains. While on the (1 0 0) grain the blisters have the 90° vertices, the blisters on the (1 1 1) grain are more round with only occasional sharpened features. Moreover, in Fig. 3a, it is clear that the coverage density of blisters on the (1 1 1) grain is smaller than that on the (1 0 0) grain. The stronger magnification in Fig. 3b shows some formation of better-defined vertices in the shape of the blisters on the (1 1 1) grain. Furthermore, we observe that (1 1 1) grain exhibits some striation at the surface that combines with the blistering. In the same image of Fig. 3b, we see well-developed blisters in clear diamond-like or rhomboidal shapes on the (1 1 0) grain as well as in Fig. 3c. A quantitative analysis will be required to obtain precise information on the density and shape of the blisters in the differently oriented grains.

The cross-sectional image of these grains shown in Fig. 3d reveals that the bubbles under the surfaces in both grains form at approximately the same depth as in the transition area (low fluence, shown in Fig. 2b), about 270–280 nm, while in this high fluence area, the bubbles are developed further in the (1 1 1) grains. We observe again the appearance of small bubbles, that are likely to be the nucleation sites of the large blisters, and as the fluence increases they develop into similar blisters as in the (1 0 0) grain. We observe that these blisters have rather hemispherical or disk-like shapes. Similarly, in Fig. 3e, we also observe the formation of blisters in the area of the high fluence, which have remarkable rhomboidal shapes of the blisters appearing at the surface. We also see the formation of small bubbles within a range of 210 and 330 nm.

*3.2. Simulation of surface protrusions grown on cu surfaces with different orientations above pressurized h bubbles*

In the experimental images shown in Figs. 2 and 3, we observe the formation of bubbles of different shapes from small rounded to large flat-faced and hemispherical ones that are more expanded in the direction of the surface. We also notice that small rounded bubbles tend to grow at an early stage of bubble formation, while the large bubbles, which usually result in surface blisters, are observed at later stages when the irradiation fluence is much higher. Although the small bubbles may interact with the surface as well, the deep positioning of these bubbles as well as the corresponding absence of visible change of surface morphology motivated us to study the effect of hydrogen accumulation in the bubble of only two shapes, the disk-like and hemispherical. These specific shapes allowed us to construct an MD model of a bubble with a large top area to achieve sufficient shear stress resulting from the effect of the gas pressure within the bubble and the presence of a nearby open surface [84,85].

We show the top views of the Cu surfaces with three different orientations, see the left column in Fig. 4 for the (1 0 0) surface (Figs. 4a and 4g), the middle column for the (1 1 1) surface (Figs. 4b and 4h) and the right column for the (1 1 0) surface (4c and 4i) after 100 ps of the relaxation run in the NVT ensemble at 600 K since a bubble of either disk-like (the upper row (4a–4c)) or hemispherical (the third row from the top (4h–4j)) shape was placed beneath it. In the second and fourth rows of Fig. 4, we show the corresponding snapshots of the structures shown in the row above, but at different times. These times were selected to show more clearly the stacking faults that were formed by Shockley partial dislocations emitted from the surface of the bubbles. The images in Figs. 4d and 4j are tilted for a better illustration of the stacking faults, while the other images are shown from the top. In these images, only the atoms with the centrosymmetry parameter more than 5 are shown for clarity. Showing only these atoms for each orientation, the formation of the surface protrusion is more visible.

In Fig. 4 we see that the large bubbles beneath the surfaces with all three orientations have resulted in protrusions with remarkably different shapes. However, all shapes of protrusions are aligned with {1 1 1} planes. This alignment and a large number of stacking faults between the bubbles and the open surface indicate that the protrusions originate from plastic deformation in Cu as a result of the interaction of hydrostatic pressure within the bubble and the open surface. Moreover, we see that the shapes of protrusions correlate with the orientation of the surface stronger than with the shape of the bubble, since regardless of the latter the shape of the protrusions are almost similar for (1 0 0), (1 1 1) and (1 1 0) surfaces (Figs. 4a and 4g, Figs. 4b and 4h, and Figs. 4c and 4i, respectively).

The most remarkable shape of square protrusions, we indeed observe on the (1 0 0) surface, which is in good agreement with most of the blisters at the surface of Fig. 2. Although none of the {1 1 1} planes is perpendicular to this surface, yet we clearly see two squares that are formed not directly above the bubble but slightly to the side from it. Unfortunately, the current size of the simulation box was too large to obtain a full protrusion of the squared shape within a reasonable computational time. To enable such a shape, we reduced the size of the simulation cell by placing the bubble closer to the surface but reducing the concentration of the H atoms ($n_{H/Vac}$ = 1.2). These results are presented in Supplemental material Section A.2. Moreover, the complex dislocation-mediated interaction of the disk-shaped bubble with the (1 0 0) surface can be viewed in the online video 100-DISK-N2.MP4.

In Fig. 4d one can see the dislocation reaction that led to the formation of the protrusion. Here we clearly see that the stacking faults (between the two Shockley partials) extended from the bubble towards the surface in four ⟨1 1 0⟩ directions on the {1 1 1} planes. The stacking faults intersect the (1 0 0) surface at the angle of 45° and the intersecting lines from four {1 1 1} planes form a square on the (1 0 0) surface as it is explained in the schematic, see Fig. 4m. Here we highlight in red the (1 0 0) surface where a protrusion is formed by the dislocations moving along four {1 1 1} planes in ⟨1 1 0⟩ directions. The dashed blue lines on the red (1 0 0) surface show the intersection lines, which form the final square shape observed in Fig. 4a. A slight offset from the position of the bubble is explained by the dislocations emitted from different random locations on the bubble surface and since they expand to the open surface at 45°, the shift of the location of some protrusion emitted from the bubble is expected. However, these shifts are unlikely to play a significant role in the shape of macroscopic protrusions, while the straight angle of the vertices of the final protrusion above the bubble will persist as the main feature of the protrusion on this surface.

A triangular protrusion appeared on the (1 1 1) surface can be seen in Fig. 4b. This differs from the close-to-circular shape of blisters seen in the experiment, see Fig. 3. The circular shape of a protrusion we observed on the surface above the bubble with the lower pressure ($n_{H/Vac}$ = 1.2), see Supplementary material Section A.2. However, we explain this circular shape of protrusion by the low gas pressure that was insufficient for the surface to yield. We note that the height of the protrusion is not uniform with the elevation that increases towards the center of the protrusion right above the under-surface bubble. This indicates the dilation of the bubble under the inner pressure as a result of the elastic response of the lattice. Consequently, when the pressure is increased and the bubble is placed deeper under the surface, we observe significant plastic deformation of the lattice in the form of dislocation network in the region between the bubble and the surface (see movie 111-DISK-N2.MP4). However, overall we can see that the triangular protrusion is the product of stacking faults growing from the bubble surface as shown in Fig. 4e. The stacking fault is forming along the three {1 1 1} planes which intersect the (1 1 1) surface in the form of a triangle (for more detail, see Fig. 4m). On the much larger scale of the experimental bubbles, one can expect the formation of a protrusion with a circular shape. This conclusion is in line with the plastic response of Cu surface to accumulated H pressure under the surface. Yet we note that the bubbles must build up higher H pressure inside before the surface with the {1 1 1} orientation can yield.

In Fig. 4c, we observe that the (1 1 0) surface yields forming a rhomboid-shaped protrusions. This result is in remarkable agreement with the experiment, see Fig. 3. Here, the stacking faults grow straight to the surface along the {1 1 1} planes that are perpendicular to the (1 1 0) surface. These planes form the angles approximately 71° and 109°, which are exactly the same angles of vertices of the obtained protrusions in our MD simulations (see Fig. 4c and 4f) and of the blisters in experiment, see Fig. 2. The angles and corresponding planes are explained in more detail in Fig. 4m. This shape of protrusion was previously observed by Pohjonen et al. [87]. For a more detailed view of the process see the online video 110-DISK-N2.MP4.

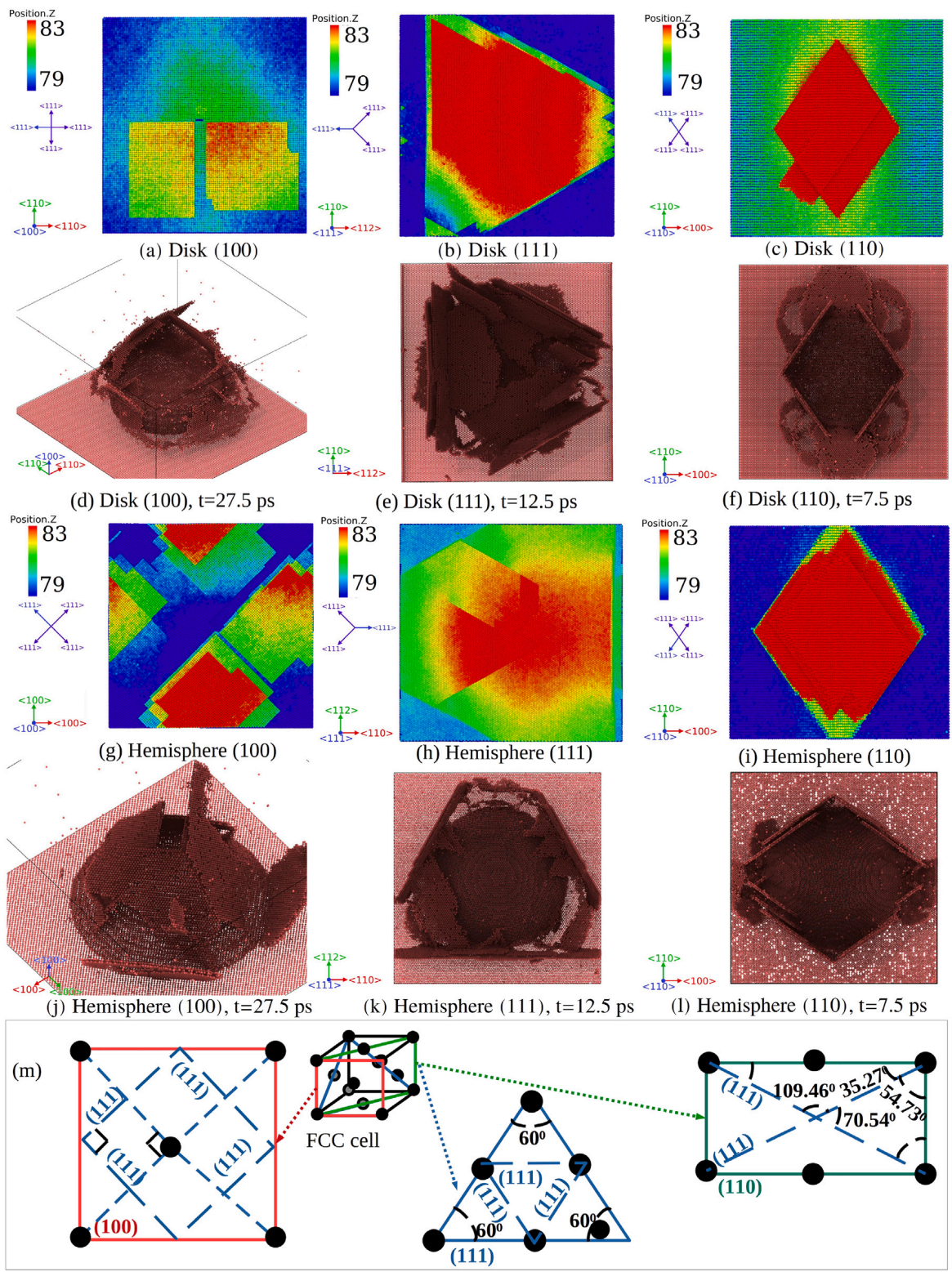

**Fig. 4.** Final configurations (after 100 ps) of surfaces for disk-shaped (a, b, c) and hemispherical-shaped bubbles (g, h, i) using (1 0 0), (1 1 1) and (1 1 0) surfaces orientation, respectively, under $n_{H/Vac}$ = 2. The directions normal to {1 1 1} planes projected onto the differently oriented surfaces are marked in blue. Atoms are colored according to their z position in Å. The lower limit corresponds to the initial level of the surface in each case, and upper limit is chosen for an easier visualization of the protrusion. Top view of cells surfaces at different times for disk-shaped (e, f, g) and hemispherical-shaped bubbles (j, k, l) oriented as (1 0 0), (1 1 1) and (1 1 0), respectively, under $n_{H/Vac}$ = 2. The view corresponding to (1 0 0) surfaces (a, d) has been tilted for improvement of clearness. Atoms located at the surface and those with centrosymmetry parameter (CSP) [86] smaller than 5 are omitted for clarity. (m) Schematic of the {1 1 1} planes projected (dashed blue lines) onto (1 0 0), (1 1 1) and (1 1 0) surfaces.

Although the overall shape of protrusions grown from the hemispherical bubble (Figs. 4g–4i and 4j–4l) are similar to the corresponding protrusions from the disk-shaped bubble, the closer comparison of the two cases reveals some differences. For instance, on the (1 0 0) surface we observe the protrusions in the form of terraces, i.e. multilayered formations that stand out from the surface much stronger than in the case of the disk-shaped bubble. It is clear that the uneven top (curvature) offers a larger area for sufficient shear stress to build up along the {1 1 1} slip planes under the hydrostatic pressure in the bubble and vicinity of the open surface facilitating emission of dislocations, compare Figs. 4a and 4g. Note the different lateral orientation of the simulation box which explains the 45° rotation of the protrusions with respect to one another on both (1 0 0) surfaces. For a detailed view of the growing dislocations see the online video 100-HEMISPHERE-N2.MP4. The same explanation holds for the hemispherical bubble under the (1 1 1) surface, see Fig. 4h. The protrusion on this surface is much less clear with many triangular shapes standing out. This is due to more frequent sites where the dislocations can be emitted. Unfortunately, the size of the cell did not allow us to continue this simulation even longer, but we observe the tendency towards a more circular shape of the protrusion compared to the disk-like bubble. To follow the complete process of dislocation emission in this case, see the online video 111-HEMISPHERE-N2.MP4.

The protrusion on the (1 1 0) surface, see Figs. 4i and 4l, appears in the rhomboidal shape, similarly to the protrusion from the disk-shaped bubble, however again, the size of this protrusion is larger. We also clearly noticed that the protrusion is also terraced as in the case of the (1 0 0) surface. This is a direct consequence of the curvature of the top of the bubble. For a detailed view of the process see movie 110-HEMISPHERE-N2.MP4.

From our MD simulations, we conclude that blisters that grow on Cu surface exposed to high fluence H irradiation can be explained by plastic deformations caused by strong H pressure that builds up in the bubbles grown under the surface. The protrusions in the form of blisters, visible on experimental surfaces, grow in our simulations via dislocations emitted from the bubble surface along the {1 1 1} planes in ⟨1 1 0⟩ directions. This explains the difference in the shapes of the blisters on the grains of different orientations.

We draw attention that the bubbles under the {1 1 0} surface resulted in a surface protrusion at the earliest time compared to the protrusions on the {1 1 1} and {1 0 0} surfaces. In fact, the latter yielded a protrusion the latest among the three orientations, see the times of the frames in the images showing the stacking faults in Fig. 4. Hence, the difference in dislocation behavior under the surfaces with different orientations cannot explain the experimental result. From the plastic deformation point of view, it is rather counter-intuitive to see the first blisters on the {1 0 0} surface, while the {1 1 1} and {1 1 0} yield much later, i.e. at much higher fluences.

Furthermore, another effect could contribute to the earlier appearance of blisters on the real {1 0 0} surface: the difference in the rate of adatom diffusion between the low-index surfaces. The activation energy of migration of an adatom on the {1 0 0} surface is ∼0.4 eV [88,89], while on the {1 1 0} surface it is 0.2–0.27 eV [90,91], and on the {1 1 1} surface only 0.04 eV [92,93]. The much faster diffusion on the {1 1 0} and {1 1 1} surfaces could facilitate the dissolution of the initially formed atomic steps before they can develop into blisters; while on the {1 0 0}, the slower diffusion makes the steps more persistent. Any difference in atom step stabilities arising from surface diffusion considerations is likewise beyond the reach of the time scales of our MD simulations and, experimentally, only large blisters are noted at the surface.

## 4. Role of crystal orientation on h accumulation under the irradiated surface

Quiros et al. [52] have reported the formation of blisters of different shapes on Al surface depending on the orientation of the grain during the exposure to $H_2$ plasma, which was also explained by plastic deformation at the formed under-surface bubbles. The shapes of the blisters that the authors observed on Al polycrystalline surface only partly resemble those that we observe in our experiments because of the different experimental conditions: $H_2$ plasma exposure compared to the ion beam with a much lower ion flux and much higher energy of tens of keV compared to hundreds of eV in plasma irradiation. We also note that the stacking fault energy in Cu lattice is lower [94] than for Al [95], which may lead to the formation of different shape protrusions, in particular, on (1 1 0) surface.

The most striking observation in our experiments emerges from the comparison of the two regions of the irradiated surface: with the low (transition area) and the high (middle of the irradiated region) ion fluences. In the transition area of our sample (see Fig. 2a) we see the well-developed blisters only on the (1 0 0)-oriented grain, but not on the (1 1 1)-oriented grain, at least, with the resolution of the SEM image. Comparing the images in Fig. 2d–2f, we see small bubbles at approximately the same depth under all three surfaces, while the large bubbles are formed only under the (1 0 0) surface. It has been suggested by Moreno and Eliezer [59] that the difference in the size of the blisters on different surfaces can be related to the channeling directions that can induce the deeper penetration of the ions. In Fig. 5a we plot the channeling map for H ions in Cu that was generated by using the MDRANGE [96] code. Indeed we see that the [1 1 0] direction exhibits the strongest channeling effect, $R_{mean} = 353 \pm 1$ nm, since this is the direction with the close-packed {1 1 1} planes along it. However, the analysis of this map does not explain the difference in blister formation between the surfaces with the [1 1 1] and [1 0 0] directions. On the contrary, the penetration of H ions in the [1 0 0] direction is slightly deeper than in the [1 1 1] one, $R_{mean} = 271 \pm 1$ nm and $R_{mean} = 255 \pm 1$ nm, respectively. Moreover, the lateral straggling cannot have a significant effect on the bubble growth, since it is fairly similar, in our simulations of the channeling map, it is $68.8 \pm 0.1$ and $69.4 \pm 0.1$ nm, respectively for (1 0 0) and (1 1 1).

We will now take a closer look at the dynamics of radiation damage under 45 keV H$^-$ ion irradiation. Since the H atoms have high binding energy to vacancies [97,98], they are known to accumulate in the vacancies, which can grow larger clusters during prolonged ion irradiation. If the H ions stop near a vacancy cluster, they can be trapped in it, forming bubbles [99]. Therefore, we now look at the distributions of vacancies in Cu and analyze their correlation with the H depth profiles during the 45 keV H$^-$ ion irradiation using the BCA code CASWIN [81]. In Fig. 5c we show the H depth profile (olive color) and the vacancies depth distribution (black color), which have maxima at similar depths with about 50 nm separation. In addition, we plot the depth distribution of the primary recoil atoms (the atoms of the target that are knocked-on directly by impacting hydrogen ion) with the energy not less than 40 eV. The energy was selected to be sufficient to create at least one vacancy in copper (according to Refs. [100,101] the threshold displacement energy for Cu is 33 eV). We see that both vacancy and primary recoil depth distributions are very similar, with a small shift towards the surface of the maximum of the vacancy depth distribution. In the CASWIN code, the crystal structure of the target material is not taken into account. Therefore, it is not possible to observe the dependence of defect distributions or depth profiles of implanted species on the crystallographic orientation of the irradiated target. For this purpose, we used the MDRANGE code, which can simulate the penetration of the H ions in a specific direction in the Cu crystal structure. For computational efficiency, the MDRANGE code does not follow the evolution of the full cascade, which would allow to obtain the vacancy depth distribution directly. Instead, the MDRANGE records the primary recoils, which can be used to assess the shape of the vacancy depth distributions in the specific directions as we saw from close comparison of the two curves in the simulations by means of the CASWIN BCA code.

The results of these calculations are shown in Fig. 5d.

In Fig. 5d we see that the ion depth profiles (open symbols) and the primary recoil profiles, i.e. the profiles of vacancy distributions (filled symbols), are strongly dependent on the crystallographic direction (see Fig. 5c, where vacancies are created a slightly closer to the surface than recoils). Comparison of both profiles, the ion depth profile and the primary recoils, for each orientation separately, reveals that only the profiles obtained in the [1 0 0] direction (red open and filled squares) have peaks located at approximately the same depth with only ∼12 nm of separation. This means that both profiles practically fully overlap, i.e. the accumulation of H in this direction occurs almost at the same depth as the created vacancies. The largest overlap of both profiles leads to the fastest bubble growth under this surface.

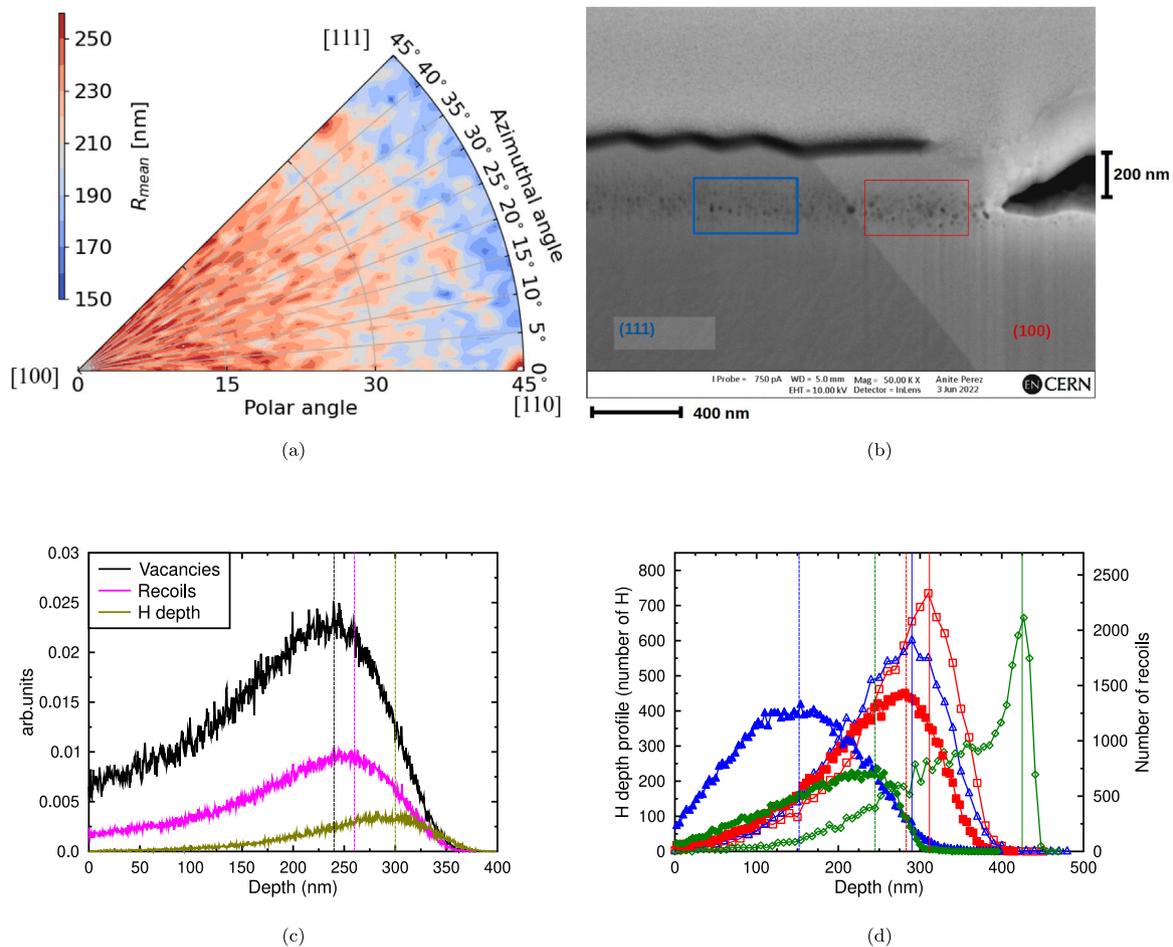

**Fig. 5.** (a) Channeling map for the 45 keV H ion irradiation of Cu calculated using the MDRANGE code [102]. The map is colored according to the mean H penetration depth (in nm, see the color bar) for each crystallographic direction identified by the pair of the polar and azimuthal angles. The main crystallographic directions are indicated at the corners of the map. The plot is generated with the polar and azimuthal angular steps of 1°, and each point is the average obtained for 100 impacts. (b) Zoom-in of grain boundary between (1 1 1) (blue) and (1 0 0) (red) grains at the transition area. The rectangular zones marked in each grain are of the same area. (c) Vacancies (black), recoil atoms with the energy above 40 eV (magenta) depth distributions as well as the H depth profile (olive) obtained by the CASWIN code, averaged over 50 000 H ion impacts. The $y$-axis shows the probability density depth distribution per ion impact. Black, magenta and olive vertical dashed lines mark the maxima of vacancy, recoil and H depth distributions, respectively. (d) H depth profile (open markers) and distribution of recoils with energy above 40 eV (filled markers) for ⟨1 0 0⟩ (square), ⟨1 1 1⟩ (triangle) and ⟨1 1 0⟩ (diamond) as a function of depth in MDRANGE averaged over 10 000 ion impacts. The recoil profiles are used to model the vacancy profile shown in (c). Vertical lines of different colors mark the maxima of the distributions of the corresponding colors. (For interpretation of the references to color in this figure legend, the reader is referred to the web version of this article.)

The ion depth profile in the [1 1 1] (blue open triangles) is similar to that in the [1 0 0] direction, but the vacancies are formed much closer to the surface. The overlap of both profiles in this direction is smaller since the peaks are separated by ∼140 nm. However, the depth of the maximum overlap is fairly close to that of the bubble growth depth in the [1 0 0] grain, the small bubbles are visible in Fig. 5b at approximately the same depth as the much larger ones that have grown in the [1 0 0] grain.

Finally, in the [1 1 0] direction, the H depth profile is the largest. We see that the separation between the peaks in this direction is larger than in the [1 1 1] direction, ∼180 nm, and the peak height is smaller when compared to the peaks of the primary recoils generated but the ions in other directions. Since the ions in this direction move along the close-packed {1 1 1} planes, i.e. in the planar channels (see the red region in the [1 1 0] direction in the channeling map in Fig. 5a), they generate fewer recoils with sufficiently high energy, which produce a lower number of vacancies. The smaller peak of the primary recoils along with the specific shape of the ion depth profile (due to channeling effect the majority of ions reach almost the maximal penetration depth) the overlap is very small and the bubble growth dynamics in this direction is remarkably slow. These simulation results explain why in experiments we see in the transition area, where the fluence is lower, only the blisters on the [1 0 0] grains and not on the other two grains, although the plastic deformation around an under-surface pressurized bubble is expected to yield faster in the grain of the ⟨1 1 0⟩ orientations, which have the slip {1 1 1} planes oriented perpendicular to the {1 1 0} surface.

Our results show that the discrepancy in penetration depths along the different directions of the incoming ions is not sufficient to explain the orientation-dependent blister formation in FCC materials [59]. Nevertheless, we observe that after high ion fluences, blistering occurs everywhere when the damage produced close to the surface is filled with hydrogen. Once hydrogen atoms occupy the created vacancies, the latter may continue growing via the loop punching mechanisms [42]. Then, after the coalescence of small bubbles into big blisters, the effect is noticed at the surface similarly as it is seen in Figs. 3d–3e and 4.

This process happens faster in (1 0 0) grains and slower in (1 1 1) and (1 1 0) grains. Consequently, the larger the difference in depth between the H depth profile and the formation of vacancies, the more fluence is needed to blister the grain.

## 5. Conclusions

In this work, we report the non-uniform blistering of a polycrystalline copper surface under the high fluence 45 keV negative hydrogen ion irradiation. The morphology analysis of the surfaces irradiated with two different ion fluences revealed that the shape of the blisters grown on different grains was different and correlated with the crystallographic orientation of the grain surface. Moreover, the whole surface of the areas irradiated to high ion fluence was densely covered with blisters fairly uniformly, while at lower fluence the coverage was highly non-uniform: well-developed blisters were seen only on some grains, while the other grains remained flat.

To obtain the insights into mechanisms of copper surface blistering, we carried out a set of molecular dynamics simulations to show how the surface of the differently oriented grains yields under the pressure built up in bubbles formed in copper during the hydrogen ion implantation. The shape of protrusions obtained from the bubbles placed under the surface with different orientations is in good agreement with the shape of blisters observed in experiments. We also were able to explain the delay for blistering on {1 1 1} and {1 1 0} oriented surfaces compared to that of {1 0 0} orientations, when we explicitly considered the crystal structure of copper. We revealed that the difference in blistering is not associated with the difference in the penetration depth of the implanted ions as it was proposed earlier but by the degree of the overlap of the hydrogen depth profile and the vacancy depth distribution. This conclusion is consistent with our experimental observation of the fastest blister growth on the grains of {1 0 0} orientations, in which the overlap of depth distributions of the implanted hydrogen atoms and the vacancies created due to ion impacts was the strongest.

**Declaration of competing interest**

The authors declare that they have no known competing financial interests or personal relationships that could have appeared to influence the work reported in this paper.

**Acknowledgments**

Computer time granted by the IT Center for Science – CSC – Finland and the Finnish Grid and Cloud Infrastructure (persistent identifier urn:nbn:fi:research-infras-2016072533) is gratefully acknowledged.

Acknowledgments to the EN-MME group at CERN for providing the time and equipment which were determinant for the analysis and imaging of the sample. The collaboration and contribution of the whole RFQ study team at CERN is also gratefully acknowledged, in particular for the sample irradiation at the RFQ test facility.

**Appendix A. Supplementary data**

Supplementary material related to this article can be found online at https://doi.org/10.1016/j.actamat.2024.119699.

A. Lopez-Cazalilla et al.Acta Materialia 266 (2024) 119699